\documentclass[aip,cha,author-year,preprint,amsmath,amssymb]{revtex4-1}

\usepackage{graphicx}
\usepackage{dcolumn}
\usepackage{bm}
\usepackage{xspace}
\usepackage{amsmath}
\draft 

\begin{document}

\newcommand{\TE}[1]{\ensuremath{T_{X \rightarrow #1}}\xspace}


\title{Inter-Scale Information Flow as a Surrogate for Downward Causation That Maintains Spiral Waves}
\author{Hiroshi Ashikaga}
\email[]{hashika1@jhmi.edu}
\homepage[]{http://www.hiroshiashikaga.org/}
\affiliation{Cardiac Arrhythmia Service, Johns Hopkins University School of Medicine, 600 N Wolfe Street, Carnegie 568, Baltimore, MD 21287}
\author{Ryan G. James}
\email[]{rgjames@ucdavis.edu}
\homepage[]{http://csc.ucdavis.edu/~rgjames/}
\affiliation{Complexity Sciences Center, Department of Physics, University of California, Davis, One Shields Avenue, Davis, CA 95616-8572}
\date{\today}

\begin{abstract}
The mechanism that maintains atrial fibrillation (AF) remains elusive. One approach to understanding and controlling the mechanism (\textit{``AF driver''}) is to quantify inter-scale information flow from macroscopic to microscopic behaviors of the cardiac system as a surrogate for the downward causation of the AF driver. We use a numerical model of a cardiac system with one of the potential AF drivers, a \textit{rotor}, the rotation center of spiral waves, and generate a renormalization group with system descriptions at multiple scales. We find that transfer entropy accurately quantifies the upward and downward information flow between microscopic and macroscopic descriptions of the cardiac system with spiral waves. Because the spatial profile of transfer entropy and intrinsic transfer entropy is identical, there are no synergistic effects in the system. We also find that inter-scale information flow significantly decreases as the description of the system becomes more macroscopic. The downward information flow is significantly smaller than the upward information flow. Lastly, we find that downward information flow from macroscopic to microscopic descriptions of the cardiac system is significantly correlated with the number of rotors, but the higher number of rotors is not necessarily associated with a higher downward information flow. This result contradicts the concept that the rotors are the AF driver, and may account for the conflicting evidence from clinical studies targeting rotors as the AF driver.

\end{abstract}

\maketitle 

\begin{quotation}
Atrial fibrillation (AF) is a powerful risk factor of stroke \citep{wolf1991atrial}, dementia \citep{ott1997atrial}, myocardial infarction \citep{soliman2014atrial}, and death \citep{benjamin1998impact}. The current standard of care for persistent AF -- continuous AF that sustains longer than 7 days \citep{january20142014} -- is invasive catheter ablation to eliminate focal triggers within the pulmonary veins that initiate AF (pulmonary vein isolation, PVI) \citep{calkins20172017}, but it remains far from curative \citep{verma2015approaches} because the mechanism that maintains AF (\textit{``driver''}) remains unknown. Improving the understanding of the AF driver to develop new and effective ablation strategies is therefore an important goal to ameliorate the suffering of 33 million patients currently affected by AF worldwide \citep{chugh2013worldwide,rahman2014global}.
\end{quotation}

\section{Introduction}

One of the striking features of the AF driver is that its downward causation \citep{campbell1974downward} from AF as a macroscopic collective behavior of the heart to microscopic behaviors of individual cardiomyocytes is clinically observable. For example, apart from the progressive structural changes, a longer duration of pacing-maintained AF results in a longer maintenance of AF after cessation of pacing \citep{wijffels1995atrial}. This phenomenon, called \textit{``AF begets AF''}, accounts for the clinical progression from paroxysmal AF to persistent AF \citep{katz1992t,yue1999molecular,bosch2003molecular,lu2008atrial}. This downward causation from macroscopic to microscopic behaviors is quantifiable as inter-scale information flow that can be used as a surrogate for the AF driver.

The objective of this study is to use a multi-scale complex systems approach to understanding downward inter-scale information flow as a surrogate for the AF driver in a cardiac system with one of the potential AF drivers, a \textit{rotor}, the rotation center of spiral waves \citep{narayan2012treatment,haissaguerre2014driver,mandapati2000stable}. To accomplish the objective, we describe spiral waves in multiple scales by generating a renormalization group from a numerical model of cardiac excitation, and quantify inter-scale information flow between macroscopic and microscopic behaviors of the system. Because rotors are considered to be a potential AF driver, we hypothesize that the higher number of rotors is associated with the higher downward inter-scale information flow.

\section{Methods}

We performed the simulation and the data analysis using Matlab R2016b (Mathworks, Inc.).

\subsection{Model of spiral waves}

We used a monodomain reaction-diffusion model that was originally derived by Fitzhugh~\citep{fitzhugh1961impulses} and Nagumo~\citep{nagumo1962active} as a simplification of the biophysically based Hodgkin-Huxley equations describing current carrying properties of nerve membranes~\citep{hodgkin1952quantitative}, which was later modified by Rogers and McCulloch~\citep{rogers1994collocation} to represent cardiac action potential. This model accurately reproduces several important properties of cardiac systems, including slowed conduction velocity, unidirectional block owing to wavefront curvature, and spiral waves.

\begin{align}
  \frac{\partial v}{\partial t} &= 0.26v(v-0.13)(1-v)-0.1vr+ I_{ex}+\nabla\cdot (D\nabla v)\\
  \frac{\partial r}{\partial t} &= 0.013(v-r)
  \label{eq:FHN02}
\end{align}

Here, $v$ is the transmembrane potential with a finite action potential duration (APD), $r$ is the recovery variable, and $I_{ex}$ is the external current~\citep{pertsov1993spiral}. $D$ is the diffusion tensor, which is a diagonal matrix whose diagonal and off-diagonal elements are 1 mm$^2$/msec and 0 mm$^2$/msec, respectively, to represent a two-dimensional (2-D) isotropic system~\citep{rogers1994collocation}. We solved the model equations using a finite difference method for spatial derivatives and explicit Euler integration for time derivatives assuming Neumann boundary conditions.

We generated 1,000 sets of a 2-D $128 \times 128$ isotropic lattice of components ($=$ 12.7 cm $\times$ 12.7 cm) by inducing spiral waves with 40 random sequential point stimulations in 40 random components of the lattice (\textit{Supporting Movie 1})\citep{ashikaga2017hidden}. In each component, we computed the time series for 10 seconds excluding the stimulation period with a time step of 0.063 msec, which was subsequently downsampled at a sampling frequency of 992 Hz to reflect realistic measurements in human clinical electrophysiology studies~\citep{fogoros2012electrophysiologic}.

We then defined the instantaneous phase $\phi (t)$ of the time series $s(t)$ in each component via construction of the analytic signal $\xi (t)$, which is a complex function of time.
\begin{equation}
\xi (t)=s(t)+is_H(t)=A(t)e^{i\phi (t)}
\label{eq:analy}
\end{equation}
Here the function $s_H(t)$ is the Hilbert transform of $s(t)$
\begin{equation}
s_H(t)=\frac{1}{\pi}\mathrm{p.v.}\int_{-\infty}^{\infty}\frac{s(\tau)}{t-\tau}d\tau
\label{eq:hilbert}
\end{equation}
where p.v. indicates that the integral is taken in the sense of the Cauchy principal value. We defined the rotor of the spiral wave as a phase singularity \citep{winfree1987time}, where the phase is undefined because all phase values converge. The phase singularity can be localized through calculation of the topological charge $n_t$ \citep{goryachev1996spiral,mermin1979topological}.
\begin{equation}
n_t= \frac{1}{2\pi}\oint_{c} \nabla\phi \cdot d \vec{l}
\label{eq:topch}
\end{equation}
where $\phi (\vec{r})$ is the local phase, and the line integral is taken over the path $\vec{l}$ on a closed curve $c$ surrounding the singularity \citep{bray2002use}.
\begin{equation}
n_t=\begin{cases}
+1 & \text{ counterclockwise rotor } \\
-1 & \text{ clockwise rotor } \\
0 & \text{ elsewhere }
\end{cases}
\label{eq:rotorcharge}
\end{equation}
In this study $|n_t|$ was used to quantify the average number of rotors over the entire time series.

\subsection{Renormalization group}

\begin{figure}[!h]
  \centering
  \includegraphics[width=\linewidth,trim={0cm 2cm 1cm 0cm},clip]{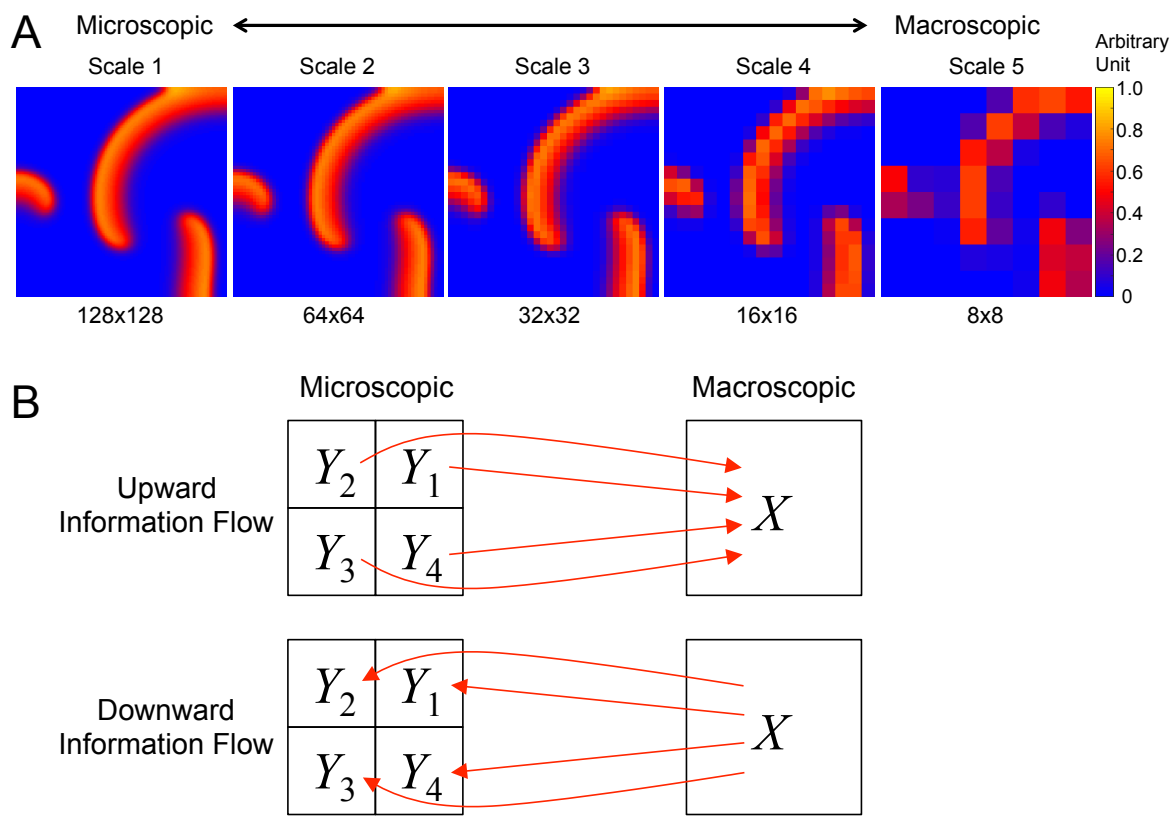}
  \caption{
    \textbf{Renormalization of a cardiac system with spiral waves.} \textit{A. Renormalization group.} This example shows a renormalization group (scale 1 through 5) of a cardiac system with three rotors. \textit{B. Upward and downward information flow.}
  }
  \label{fig:concept}
\end{figure}

Using the 2-D $128 \times 128$ isotropic lattice of components as the original microscopic description of the system (\textit{scale 1}, $\Delta x=$ 0.99 mm), we generate a renormalization group of the system by a series of transformation including coarse-graining and length rescaling. For example, we average over the values of the excitation variable $v$ at each time point in the block of 2 $\times$ 2 immediately adjacent components of the system at scale 1, and assign the mean value of the excitation variable $v$ to the corresponding site in the system at scale 2 ($64 \times 64$ lattice, $\Delta x=$ 1.98 mm) \citep{mccomb2004renormalization}. In a serial fashion we generate the macroscopic description of the system at scale 3 ($32 \times 32$ lattice, $\Delta x=$ 3.96 mm), scale 4 ($16 \times 16$ lattice, $\Delta x=$ 7.92 mm), and scale 5 ($8 \times 8$ lattice, $\Delta x=$ 15.84 mm) (Figure~\ref{fig:concept}A, Supporting Movie 2).

\subsection{Information flow between scales}

We treat each component on the lattice as a time-series process $X$. \textit{Shannon entropy} $H$ of each time-series process $X$ is
\begin{eqnarray}
H(X)&=&-\sum_{x}p(x)\log_{2}p(x)
\label{eq:entropy}
\end{eqnarray}
where $p(x)$ denotes the probability density function of the time series generated by $X$. \textit{Mutual information} $I(X;Y)$ is a measure of the reduction in uncertainty of the time-series process $X$ due to the information gained from knowing the time-series process $Y$.

\begin{eqnarray}
I(X;Y)&=&H(X)+H(Y)-H(X,Y)\\
&=&\sum_{x,y}p(x,y)\log_{2}\frac{p(x,y)}{p(x)p(y)}
\label{eq:mi}
\end{eqnarray}
where $p(x,y)$ and $H(X,Y)$ denote the joint probability density function and the joint entropy of $X$ and $Y$, respectively. When in the presence of a third variable $Z$, we can quantify the additional reduction in uncertainty about variable $Y$ given $X$, after already having been given $Z$ using \textit{conditional mutual information}.

\begin{eqnarray}
I(X;Y|Z)&=&H(Y|Z)-H(Y|X,Z)\\
&=&\sum_{x,y,z}p(x,y|z)\log_{2}\frac{p(x,y|z)}{p(x|z)p(y|z)}
\label{eq:cmi}
\end{eqnarray}

\textit{Transfer entropy} \citep{schreiber2000measuring} from a process $X$ to another process $Y$ is the amount of uncertainty reduced in future values of $Y$ by knowing the past values of $X$, given past values of $Y$.
\begin{align}
  \TE{Y}
  &= I(x^k_t;y_{t+1}|y^l_t)\\
  &= H(y_{t+1}|y^l_t) - H(y_{t+1}|y^l_t,x^k_t)\\
  &= \sum p(y_{t+1},y^l_t,x^k_t) \log_2 \frac{p(y_{t+1}|y^l_t,x^k_t)}{p(y_{t+1}|y_t^l)}
  \label{eq:Txy}
\end{align}
$k$ and $l$ denote the length of time series in the processes $X$ and $Y$, respectively.
\begin{align}
  x^k_t &= (x_t,x_{t-1},...,x_{t-k+1}) \\
  y^l_t &= (y_t,y_{t-1},...,y_{t-l+1})
  \label{eq:xkyl}
\end{align}
In this study we define $k=l=1$. We calculate transfer entropy in continuous time series of the transmembrane potential $v$, and also in binary time series with 1 when excited (during APD$_{90}$) and 0 when resting \citep{ashikaga2015modelling} for comparison.

We evaluate upward and downward information flow between scales on a 2-D lattice (Figure~\ref{fig:concept}B). We define the upward information flow as transfer entropy from a microscopic component to the corresponding component on the next macroscopic scale. Because a block of 2 $\times$ 2 immediately adjacent microscopic components corresponds to one component on the next macroscopic scale, a total of four upward information flow values are calculated per macroscopic component. Likewise, we define the downward information flow as transfer entropy from a macroscopic component to the corresponding block of 2 $\times$ 2 immediately adjacent components on the next microscopic scale. Similar to the the upward information flow, a total of four downward information flow values are calculated per macroscopic component.

Transfer entropy, due to its formulation as a conditional mutual information, captures both direct information transfer between two time series as well as synergistic interactions between the two \citep{bossomaierintroduction}. To exclude such synergistic effects from the conditional mutual information, Maurer and Wolf proposed \textit{intrinsic mutual information} \citep{maurer1997intrinsic,christandl2003property}, which destroys synergistic interactions by minimizing conditional mutual information over all local transformations of $Z$.

\begin{align}
  I(X;Y \downarrow Z) = \min_{p(\overline{z}|z)} I(X;Y|\overline{Z})
\end{align}

Because the minimization is taken over all possible corruptions of $Z$, any synergistic interactions between $X$, $Y$, and $Z$ are eleminated. Nevertheless, dependence between $X$ and $Y$ are preserved because $Z$ is corrupted locally; that is, the minimization is over $p(\overline{z}|z)$, not $p(\overline{z}|x,y,z)$. Likewise, \textit{intrinsic transfer entropy} \citep{james2017ite}, captures the dependence between $x^k_t$ and $y_{t+1}$ excluding synergistic interactions between $X$ and $Y$.

\begin{align}
IT_{X \rightarrow Y}
&= I(x^k_t;y_{t+1}\downarrow y^l_t)\\
&= \min_{p(\overline{y^l_t}|y^l_t)} I(x^k_t;y_{t+1}|\overline{y^l_t})
\end{align}

If transfer entropy and intrinsic transfer entropy are identical for two given time series, one can conclude that synergistic effects do not exist in the system and transfer entropy captures direct information transfer alone. We calculate intrinsic transfer entropy in binary time series with 1 when excited (during APD$_{90}$) and 0 when resting \citep{ashikaga2015modelling} for comparison. We compute the intrinsic transfer entropy using binary rather than continuous time series because performing the optimization over all possible local corruptions of $Z$ in the continuous case is vastly more involved and is beyond the scope of this study.

For continuous time series we use the nearest-neighbor based Kraskov-St{\"o}gbauer-Grassberger (KSG) estimator \citep{kraskov2004estimating} extended to transfer entropy \citep{gomez2015assessing} that is implemented in an open-source library (Java Information Dynamics Toolkit; \url{http://jlizier.github.io/jidt/}) \citep{lizier2014jidt}. For binary time series we use another open-source library (Discrete Information Tool; \url{https://github.com/dit/dit}) to calculate transfer entropy and intrinsic transfer entropy. We use custom Matlab and Python code to adapt the library to calculate information metrics in our data set. We adopt the standard convention that $0 \cdot \log_2{0} = 0$. We use natural logarithms with a base of $e$ and the resulting information measures will have units of the natural unit of information (nat).

\section{Results}

\subsection{Evaluation of transfer entropy to quantify inter-scale information flow}

\begin{figure}[!h]
  \centering
  \includegraphics[width=\linewidth,trim={1cm 1cm 1cm 0cm},clip]{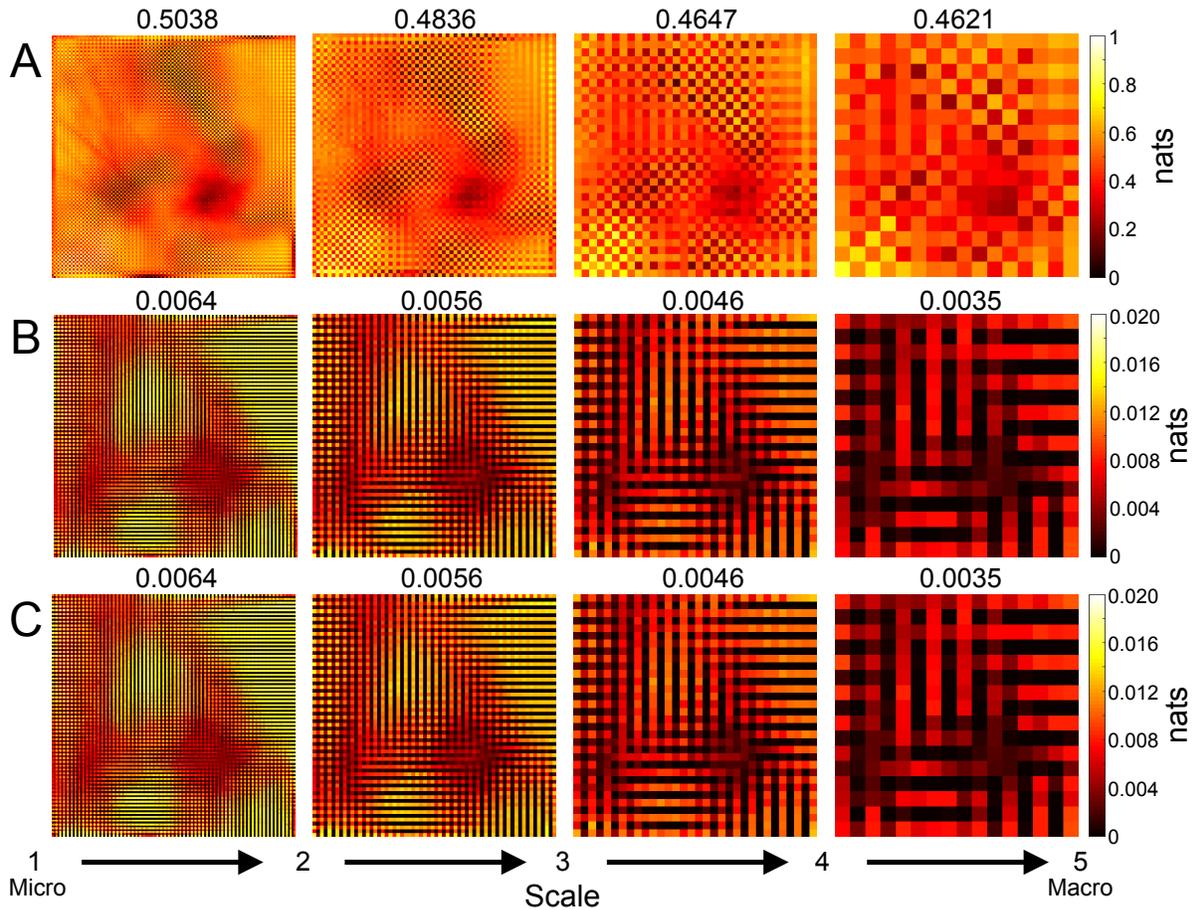}
  \caption{
    \textbf{Upward information flow in a representative spiral wave data set.} The columns represent the scale from microscopic to macroscopic description of the system (scale 1 through 5) and the rows represent different methods of transfer entropy calculation, including the KSG estimator for continuous time series ($A$), the histogram for binary time series ($B$), and intrinsic transfer entropy for binary time series ($C$). The mean transfer entropy is shown on top of each lattice. The unit is nats.
  }
  \label{fig:upwardTE}
\end{figure}

First we evaluate whether transfer entropy accurately captures inter-scale information flow without synergistic effects in a representative spiral wave data set. Overall, upward inter-scale transfer entropy is regionally heterogeneous, and the mean transfer entropy over the lattice decreases from a microscopic scale to a macroscopic scale (Fig.\ref{fig:upwardTE}). The spatial profile of inter-scale transfer entropy is quantitatively different but qualitatively similar between continuous (Fig.\ref{fig:upwardTE}A) and binary time series (Fig.\ref{fig:upwardTE}B). Importantly, the spatial profile of inter-scale transfer entropy (Fig.\ref{fig:upwardTE}B) and inter-scale intrinsic transfer entropy (Fig.\ref{fig:upwardTE}C) is identical. This indicates that there are no synergistic effects in the system, and thus transfer entropy is an appropriate metric to quantify upward inter-scale information flow from microscopic to macroscopic descriptions of the system.

\begin{figure}[!h]
  \centering
  \includegraphics[width=\linewidth,trim={1cm 1cm 1cm 0cm},clip]{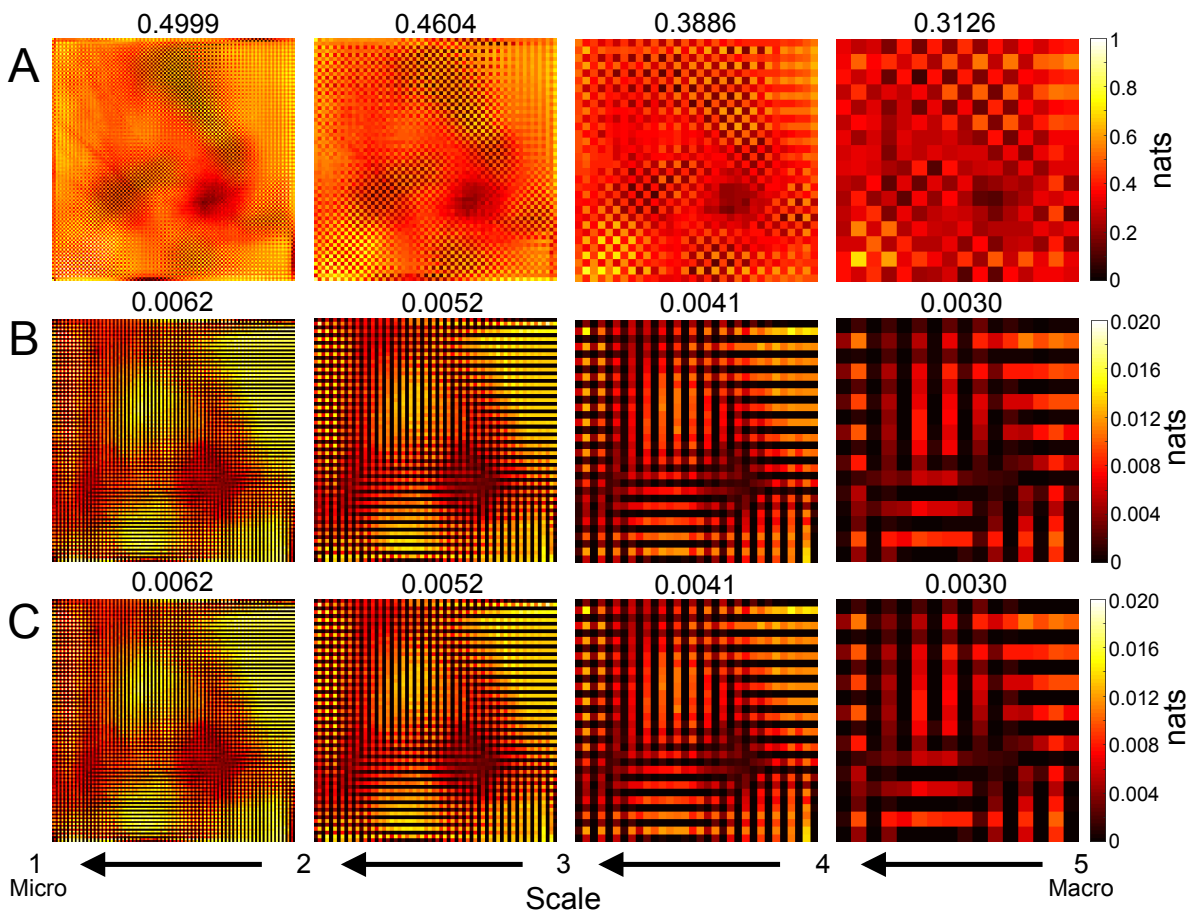}
  \caption{
    \textbf{Downward information flow in a representative spiral wave data set.} The columns represent the scale from microscopic to macroscopic descriptions of the system (scale 1 through 5) and the rows represent different methods of transfer entropy calculation, including the KSG estimator for continuous time series ($A$), the histogram for binary time series ($B$), and intrinsic transfer entropy for binary time series ($C$). The mean transfer entropy is shown on top of each lattice. The unit is nats.
  }
  \label{fig:downwardTE}
\end{figure}

Using the same spiral wave data set, we evaluate the validity of downward inter-scale transfer entropy as information flow. The spatial profile of downward inter-scale transfer entropy of the continuous (Fig.\ref{fig:downwardTE}A) and binary time series (Fig.\ref{fig:downwardTE}B) is similar to those of upward inter-scale transfer entropy both qualitatively and quantitatively. However, the mean transfer entropy over the lattice of downward inter-scale transfer entropy is smaller than that of upward inter-scale transfer entropy. Similar to upward inter-scale transfer entropy, the spatial profile of downward inter-scale transfer entropy (Fig.\ref{fig:downwardTE}B) and downward inter-scale intrinsic transfer entropy (Fig.\ref{fig:downwardTE}C) is identical. This indicates that there are no synergistic effects in the system, and thus transfer entropy is an appropriate metric to quantify downward inter-scale information flow from macroscopic to microscopic descriptions of the system.

\subsection{Inter-scale information flow in aggregate data sets}

\begin{figure}[!h]
  \centering
  \includegraphics[width=\linewidth,trim={1cm 7cm 1cm 0cm},clip]{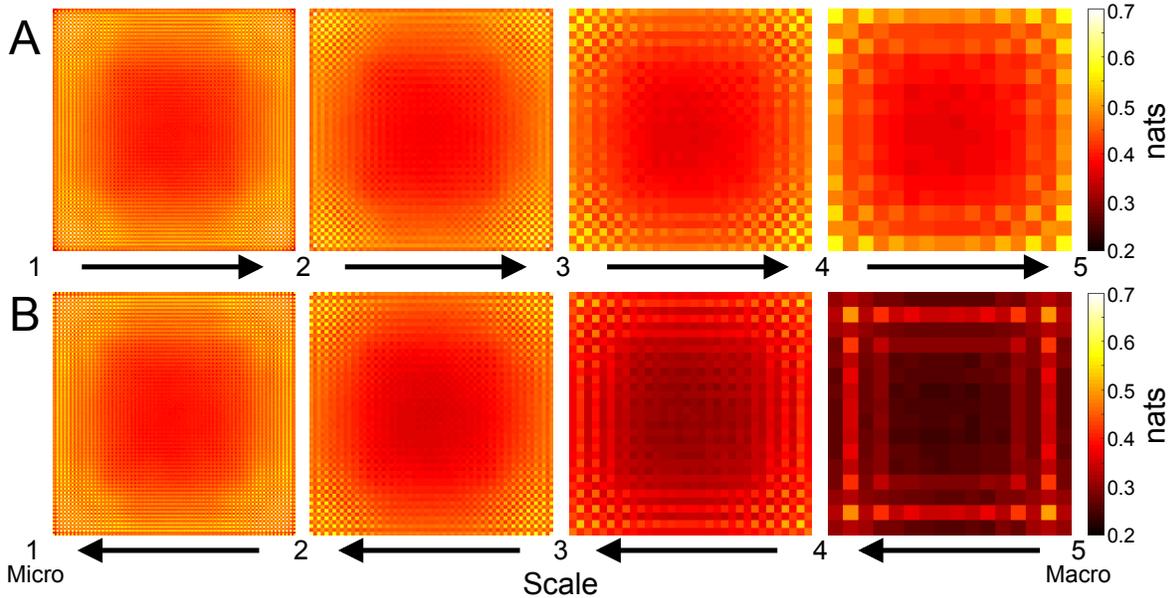}
  \caption{
    \textbf{Inter-scale information flow in aggregate data sets.} The value of each component represents the mean transfer entropy of 1,000 data sets with different numbers of rotors. The columns represent the scale from microscopic to macroscopic descriptions of the system (scale 1 through 5) and the rows represent upward (\textit{A}) and downward transfer entropy (\textit{B}) calculated by the KSG estimator for continuous time series. The unit is nats.
  }
  \label{fig:meanTE}
\end{figure}

\begin{figure}[!h]
  \centering
  \includegraphics[width=\linewidth,trim={1cm 4cm 3cm 0cm},clip]{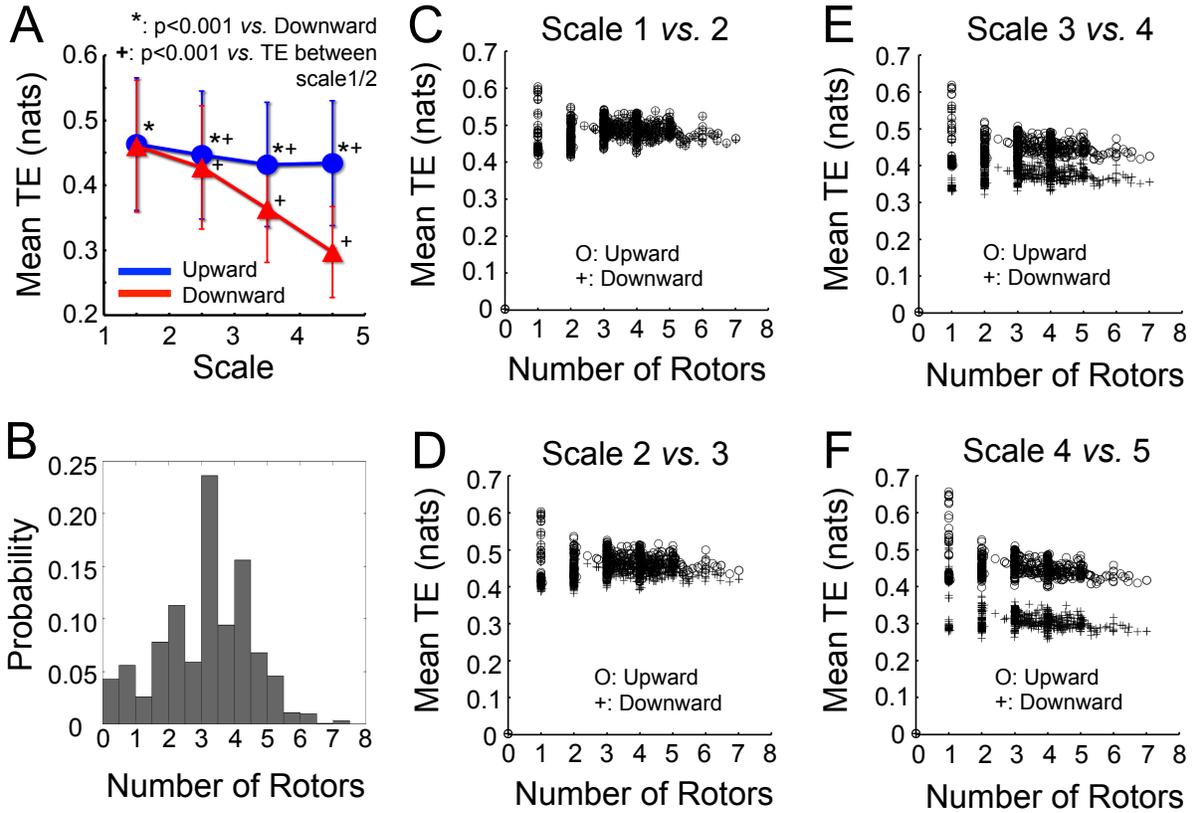}
  \caption{
    \textbf{Inter-scale information flow and number of rotors in aggregate data sets.} \textit{A. Mean transfer entropy (TE) and scale} *: p$<$0.001 \textit{vs.} downward TE, +: p$<$0.001 \textit{vs.} TE between scale 1 and 2. \textit{B. Probability distribution of number of rotors}. The number of rotors ranges from 0 to 7. The number of rotors and mean TE between scale 1 and 2 (\textit{C}), between scale 2 and 3 (\textit{D}), between scale 3 and 4 (\textit{E}), and between scale 4 and 5 (\textit{F}). Upward information flow ($\bigcirc$); Downward information flow (+). The unit is nats.
  }
  \label{fig:scatter}
\end{figure}

Next we quantify upward and downward inter-scale information flow between microscopic and macroscopic descriptions of the system in aggregate data of 1,000 sets with different numbers of rotors in continuous time series (Fig.\ref{fig:meanTE}). The spatial profile of the mean transfer entropy in the aggregate data sets shows that the regional heterogeneity is almost completely smoothed out, leaving higher transfer entropy values only in the borders of the  lattice. Overall, transfer entropy of downward inter-scale transfer entropy is smaller than that of upward inter-scale transfer entropy, which is consistent with a specific case (Fig.\ref{fig:upwardTE}A,\ref{fig:downwardTE}A). The mean of both upward and downward inter-scale transfer entropy significantly decreases from microscopic to macroscopic descriptions of the system, and downward inter-scale transfer entropy is significantly smaller than upward inter-scale transfer entropy (Fig.\ref{fig:scatter}A).

\subsection{Impact of rotors on inter-scale information flow}

Lastly we evaluate the impact of rotors on inter-scale information flow. The number of rotors ranges from 0 to 7, with a median of 3 (Fig.\ref{fig:scatter}B), and we use data sets with at least one rotor. Between any scales, the number of rotors is significantly correlated with both the mean upward and downward transfer entropy, but the correlaton coefficient is relatively small (Fig.\ref{fig:scatter}C,Table~\ref{tab:table1}). For example, between scale 1 and 2, the correlaton coefficient of the mean upward and downward transfer entropy is 0.2690 and 0.2464, respectively. This indicates that a higher number of rotors is significantly associated with a higher value of both upward and downward transfer entropy, but the scatter is large. As the scale goes up the correlaton coefficient of the mean upward and downward transfer entropy goes even lower, and becomes negative between scale 4 and 5. This indicates that on higher scales a higher number of rotors is significantly associated with a lower value of both upward and downward transfer entropy.

\begin{table}
\centering
\caption{
\textbf{Correlation between mean transfer entopy and number of rotors} r, correlation coefficient }
\begin{tabular}{| c | c | c | c | c | c |}
\hline
 &  & Scale 1 \textit{vs.} 2 & Scale 2 \textit{vs.} 3 & Scale 3 \textit{vs.} 4 & Scale 4 \textit{vs.} 5\\ \hline
Upward & r & 0.2690 & 0.2183 & 0.1173 & -0.1715\\
       & p & $<$0.001	& $<$0.001 & $<$0.001 & $<$0.001\\
Downward & r & 0.2464 &	0.1499 & 0.090 & -0.1448\\
       & p & $<$0.001	& $<$0.001	 & 0.0068 & $<$0.001\\ \hline
\end{tabular}
\label{tab:table1}
\end{table}

\section{Discussion}

\subsection{Main findings}

First, we find that transfer entropy accurately quantifies the upward and downward information flow between microscopic and macroscopic descriptions of the cardiac system with spiral waves. Because the spatial profile of transfer entropy and intrinsic transfer entropy is identical, there are no synergistic effects in the system.

Second, we find that inter-scale information flow significantly decreases as the description of the system becomes more macroscopic. The downward information flow is significantly smaller than the upward information flow.

Lastly, we find that inter-scale information flow is significantly correlated with the number of rotors, but the scatter is large. In addition, the higher number of rotors is not necessarily associated with a higher downward information flow. At microscopic scales, a higher number of rotors is associated with a higher downward information flow. As the system description becomes more macroscopic, the higher number of rotors is associated with a lower downward information flow.

\subsection{Multi-scale complex systems approach to atrial fibrillation}

Our approach contains several innovative aspects. First, we use a multi-scale complex systems approach to AF to characterize inter-scale information flow. Until now, scientific investigation of AF has focused on describing the behavior of microscopic components at specific scales, including genes, proteins, ion channels, cells, and tissues \citep{Lip:2016aa}. However, the heart is a multi-scale complex system consisting of multitude of cells, including five billion autonomous cardiomyocytes \citep{kapoor2013direct}, with simple rules of operation and minimal central control \citep{ashikaga2015modelling}. Therefore, an opportunity exists for understanding and controlling AF at multiple scales. This study evaluates the inter-scale information flow from macroscopic to microscopic scales as a surrogate for the downward causation of the AF driver.

Second, we utilize the renormalization group where we apply iterated coarse-graining and rescaling \citep{kadanoff1993scaling} to the microscopic description of the cardiac system to construct a series of robust and minimal
macroscopic descriptions. The renormalization group offers a generalized and systematic multi-scale analysis to derive macro-scale behaviors and provides a systematic means of extracting macro-scale features and reducing the number of degrees of freedom. The renormalization group is particularly powerful when analyzing phase transitions where long-range correlations build up a singular macroscopic behavior \citep{ashikaga2017locating}. This is an exceedingly innovative approach, because it is a completely opposite concept of a conventional and common belief that a detailed, high-resolution modeling with near-complete description of microscopic behaviors with infinite degrees of freedom is required to understand the macroscopic behavior of the cardiac system.

Third, we use intrinsic transfer entropy to evaluate the synergistic effects between microscopic and macroscopic descriptions of the cardiac system. Transfer entropy is a form of conditional mutual information, which captures both intrinsic dependence between variables as well as conditional dependence. Rather than utilizing conditional mutual information, intrinsic transfer entropy uses intrinsic mutual information from information-theoretic cryptography. This provides for the first time a concrete method of separately quantifying intrinsic information flow from conditional information flow. In this study we demonstrate that transfer entropy and intrinsic transfer entropy between scales are identical, therefore there are no synergistic effects in the system and transfer entropy captures direct information transfer alone.

\subsection{Clinical implications}

Successful ablation requires targeted perturbation of the mechanism that maintains arrhythmia. Earlier strategies of persistent AF ablation, such as posterior wall isolation \citep{sanders2007complete,chen2008treatment}, a ``stepwise'' approach \citep{jais2004technique,fassini2005left,scherr2014five}, and the Cox-Maze procedure \citep{damiano2011cox}, focused on segmenting the atria with linear ablation lesions to reduce the mass of contiguous atrial tissue below a ``critical mass'' needed to sustain fibrillation within the myocardium \citep{mcwilliam1887fibrillar,garrey1914nature}. However, a prospective randomized study showed that empiric linear lesions in addition to PVI do not improve the recurrence rates compared with PVI alone \citep{verma2015approaches}. More recent strategies include targeting the rotor of spiral waves as a potential AF driver, with promising results from early clinical studies  \citep{narayan2012treatment,miller2014initial,narayan2014ablation,haissaguerre2014driver}.
In this study we find that the higher number of rotors is not necessarily associated with a higher downward information flow. If rotors are the AF driver, the relationship between the number of rotors and downward information flow should be scale-invariant. Therefore, our results cast a shadow of doubt on the concept of rotors as the AF driver. This may explain why more recent studies targeting rotors showed disappointing results \citep{benharash2015quantitative,gianni2016acute,berntsen2016focal,share2014clinical}.

\subsection{Limitations}

We used a modified Fitzhugh-Nagumo model, which is a relatively simple model of excitable media, with a homogeneous and isotropic lattice. It is possible that our findings may not directly be extrapolated to a more realistic cardiac system with tissue heterogeneity and anisotropy. However, the information-theoretic metrics in this study are independent of any specific trajectory of each rotor. Therefore, our approach is applicable to any other cardiac system.

\subsection{Conclusions}

Downward information flow from macroscopic to microscopic descriptions of the cardiac system is significantly correlated with the number of rotors, but the higher number of rotors is not necessarily associated with a higher downward information flow. This result contradicts the concept that the rotors are the AF driver, and may account for the conflicting evidence from clinical outcome studies targeting rotors as the AF driver.

\section*{Supplementary material}

\paragraph*{Supporting Movie 1.}
\label{sample_stim_movie}
\textit{Random sequential point stimulations.} We induce spiral waves by introducing 40 random sequential point stimulations in 40 random components of the lattice. In this example, random sequential point stimulations induce five spiral waves.

\paragraph*{Supporting Movie 2.}
\label{renorm}
\textit{Renormalzation group.} The movie shows a renormalization group of the cardiac system with three spiral waves by a series of transformation including coarse-graining and length rescaling (scale 1 through 5).

\begin{acknowledgments}
This work was supported by the Fondation Leducq Transatlantic Network of Excellence (to H.A.), and the U. S. Army Research Laboratory and the
U. S. Army Research Office under contracts W911NF-13-1-0390 and
W911NF-13-1-0340, and the UC Davis Intel Parallel Computing Center (to R.G.J.).
\end{acknowledgments}

\bibliography{renorm}

\end{document}